\documentclass{elsart3p}
\usepackage[T1]{fontenc}
\usepackage{amsmath}
\usepackage{amssymb}
\usepackage{latexsym}
\usepackage{graphicx}

\begin{document}
\begin{frontmatter}

\title{When did cosmic acceleration start? How fast was the transition?}
\author{\'Emille E. O. Ishida},
\ead{emille@if.ufrj.br}
\author{Ribamar R. R. Reis},
\ead{ribamar@if.ufrj.br}
\author{Alan V. Toribio},
\ead{alan@if.ufj.br}
\author{Ioav Waga}
\ead{ioav@if.ufrj.br}

\address{Universidade Federal do Rio de Janeiro, Instituto de
F\'{\i}sica, CEP 21941-972, Rio de Janeiro, RJ, Brazil}

\begin{abstract}
Cosmic acceleration is investigated through a kink-like expression
for the deceleration parameter ($q$). The new parametrization
depends on the initial ($q_i$) and final ($q_f$) values of $q$, on
the redshift of the transition from deceleration to acceleration
($z_{t}$) and the width of such transition ($\tau $). We show that
although supernovae (SN) observations (\textit{Gold182} and SNLS
data samples) indicate, at high confidence, that a transition
occurred in the past ($z_{t}>0$) they do not, by themselves, impose
strong constraints on the maximum value of $z_{t}$. However, when we
combine SN with the measurements of the ratio between the comoving
distance to the last scattering surface and the SDSS+2dfGRS BAO
distance scale ($S_{k}/D_{v}$) we obtain, at $95.4\%$ confidence
level, $z_{t}=0.84\pm _{0.17}^{0.13}$ and $\tau =0.51\pm
_{0.17}^{0.23}$ for ($S_{k}/D_{v}$+\textit{Gold182}), and $%
z_{t}=0.88\pm _{0.10}^{0.12}$ and $\tau =0.35\pm _{0.10}^{0.12}$ for ($%
S_{k}/D_{v}$ + SNLS), assuming $q_i=0.5$ and $q_f=-1$. We
also analyze the general case, $q_f\in(-\infty,0)$ finding the constraints that the combined tests ($%
S_{k}/D_{v}$ + SNLS) impose on the present value of the deceleration
parameter ($q_0$).
\end{abstract}

\end{frontmatter}

\section{Introduction}

Since the discovery of the accelerated expansion of the universe in
1998 \cite{riess98, perlm99}, considerable effort in cosmology has
been devoted to determine the source of this acceleration. The two
most common possibilities  discussed in the literature are: the
existence of an exotic component with sufficiently negative pressure
(dark energy) and proper modifications of general relativity at
cosmological scales (for recent reviews see \cite{review}).

One way of making progress in determining the cosmic expansion
history is through a model by model analysis. Another is to carry
out a phenomenological analysis with the use of different
parameterizations of the dark energy equation of state
\cite{eosparam}, the Hubble parameter \cite{hubble} or the dark
energy density \cite{wang01}. This procedure may provide interesting
pieces of information, but in general a parametrization assumes the
existence of dark matter and dark energy as different substances
(barring a few exceptions no interaction in the dark sector is
considered) and general relativity is in most cases assumed. In this
framework, an important question regards the number of parameters
necessary to get reliable conclusions. If too many are used, the
allowed region in the parameter space could be so large that it
would not be possible to get firm conclusions \cite{linder05}.
Otherwise, if not enough parameters are used, the obtained results
may be strongly dependent on the particular parametrization choice
and misleading conclusions could be reached \cite{bassett04}. The
strategy we follow here is to use a large (four) number of
parameters in order to be quite general, but, based on physical
arguments, fix two of them from the start. We then relax the
condition on one of the fixed parameters and obtain the confidence
surface on the other three.

In this work we are mainly interested in the following questions:
what is the redshift of the transition from decelerated to
accelerated expansion? How fast was it? We investigate these by
introducing a new parametrization for the deceleration parameter
($q$) that depends on four parameters: the initial ($q_i$) and final
($q_f$) values of $q$, the redshift of the transition from
deceleration to acceleration ($z_{t}$) and a quantity related to the
width in redshift of such transition ($\tau $). With this
formulation we aim to answer the above questions with the minimum
amount of assumptions about the dark sector and  the fundamental
gravitation theory.

This paper is organized as follows: in Section 2, we present the new
$q$ parametrization and discuss some of its properties. In Section 3
the outcomes of the confrontation of this parametrization with two
supernovae samples, the new \textit{Gold182} SNe Ia from \cite{gold}
and the first year data set of the Supernova Legacy Survey (SNLS)
\cite{snls}, are obtained (first assuming $q_f=-1$ and $q_i=1/2$).
We show that current supernovae observations alone are not able to
satisfactorily constrain the transition redshift. To break the SN Ia
degeneracy, we combine this observable with the ratio of the
comoving distance to the last scaterring surface
($S_{k}(z_{ls}=1098)$) to the baryon acoustic oscillations (BAO)
distance scale ($D_{v}(z)$) at $ z_{BAO}=0.2$ and $z_{BAO}=0.35$ as
estimated in \cite{percival07}. We show that the $\Lambda$CDM model
is within the region allowed by SNLS+$S_k/D_v$ results but is
excluded for \textit{Gold182}+$S_k/D_v$ data, at $95\%$ confidence
level. We then discuss the broader case with arbitrary $q_f$
exhibiting the $95\%$ confidence surface in the parameter space
$(z_t,\tau,q_0)$ ($q_0$ is the present value of $q$), obtained using
the SNLS+ $S_{k}/D_{v}$ data. Our conclusions are presented in
Section 4.

\section{The Model}

At large scales, it is a good approximation to consider a spatially
homogeneous and isotropic universe. With this assumption we are lead
to the Friedman-Robertson-Walker metric:
\begin{equation}
ds^{2}=dt^{2}-a(t)^2\left[ \frac{dr^{2}}{1-kr^{2}}+r^{2}d\Omega^2
\right] ,
\end{equation}%
where $a(t)$ is the scale factor and $k=-1,0,+1$ characterizes the curvature
of the spatial sections of space-time. From now on we will assume a flat
universe ($k=0$), which is in agreement with CMB results \cite{spergel}.

In terms of the Hubble parameter ($H\equiv \frac{\dot{a}}{a}$), the
deceleration parameter can be written as:
\begin{equation}
q=-\frac{\ddot{a}}{aH^{2}}=\frac{d}{dt}\left( \frac{1}{H}\right) -1.
\end{equation}%
Therefore,
\begin{equation}
H=H_{0}\exp {\left[ \int_{0}^{z}(q(\tilde{z})+1)d\ln
{(1+\tilde{z})}\right] }. \label{H}
\end{equation}

\begin{figure*}[tbp]
\begin{center}
\includegraphics[width=16cm,
  angle=0]{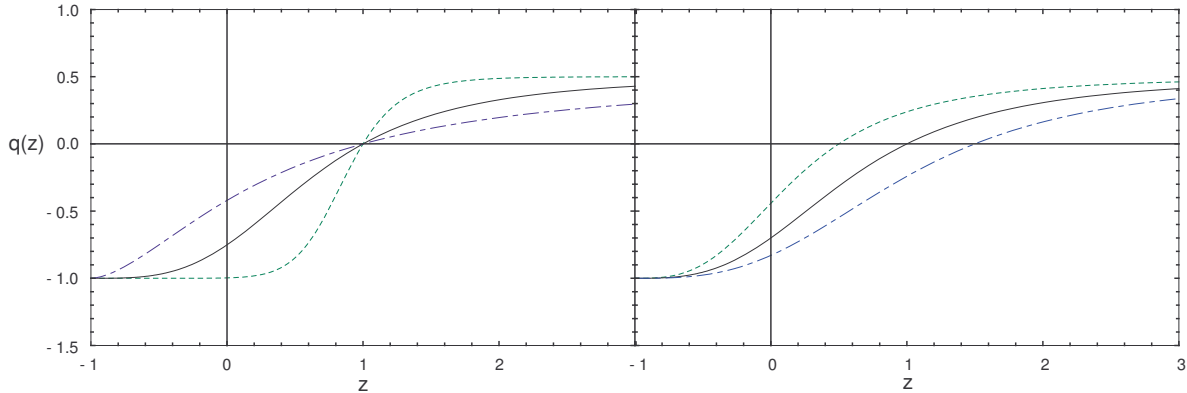}
\end{center}
\caption{{\protect {Influence of parameters $z_t$ and $\tau$ in the
functional form of the deceleration parameter for the special case
$q_i=0.5$ and $q_f=-1$. \textbf{Left} - $q(z)$ for $z_t=1.0$ and
$\tau$=0.1 (green dashed), 0.3 (full black) and 0.5 (blue
dot-dashed). \textbf{Right} - $q(z)$ for $\tau=0.3$ and $z_t=$ 0.5
(green dashed), 1.0 (full black) and 1.5 (blue dot-dashed).}}}
\label{fig3}
\end{figure*}

In this work, we propose the following phenomenological functional
dependence with redshift for the deceleration parameter:
\begin{equation}
q(z)\equiv q_{f}+\frac{(q_{i}-q_{f})}{1-\frac{q_{i}}{q_{f}}\left( \frac{%
1+z_{t}}{1+z}\right) ^{1/\tau }},  \label{novoq}
\end{equation}%
where $q_{i}>0$ (deceleration) and $q_{f}<0$ (acceleration) are the initial $%
(z\gg z_{t})$ and final $(z=-1)$ values of the deceleration parameter,
respectively. The parameter $z_{t}$ denotes the redshift of the transition ($%
q(z_{t})=0$) and $\tau >0$ is associated with the width of the
transition. It is related to the derivative of $q$ with respect to
the redshift at $z=z_{t}$. More precisely,
\begin{equation}
\tau ^{-1}=\left(\frac{1}{q_{i}}-\frac{1}{q_{f}}\right)\left[\frac{dq(z)}{d\ln (1+z)}%
\right]_{z=z_{t}}.  \label{tau}
\end{equation}%
The influence of parameters $z_t$ and $\tau$ are demonstrated in
Fig. (\ref{fig3}).

Expression (\ref{novoq}) is similar in spirit to the one suggested in \cite%
{bassett} (see also \cite{linder05,bassett04,kink}), but here we
parametrize $q(z)$ instead of $w(z)$. One of the advantages of using
the above kink-like parametrization for the deceleration parameter
is that $z_{t}$ has a very clear physical meaning. Different
physical aspects and parameterizations of $q(z)$ were also
investigated in \cite{qparam}.
With the above definition, equation (\ref{H}) is now integrated to
give,
\begin{eqnarray}
\left( \frac{H(z)}{H_{0}}\right) ^{2}&=&\left( 1+z\right)
^{2(1+q_{i})} \nonumber \\
& & \times \left( \frac{q_{i}\left( \frac{1+z_{t}}{1+z}\right)
^{1/\tau }-q_{f}}{q_{i}\left( 1+z_{t}\right) ^{1/\tau
}-q_{f}}\right) ^{2\tau (q_{i}-q_{f})}. \label{hq0}
\end{eqnarray}
We now define an effective matter density parameter ($\Omega
_{m\infty }$) as
\begin{equation}
\Omega _{m\infty } \equiv \lim_{z\rightarrow \infty }\left( \frac{H(z)}{%
H_{0}}\right) ^{2}\left( 1+z\right) ^{-2(1+q_i)},
\end{equation}%
where the limit should be understood as $z>>z_t$.

In most (and simplest) scenarios, in order to form large scale
structures, the universe passes trough a kind of matter dominated
phase such that, at early times (but after radiation domination),
$H^{2}\propto (1+z)^{3}$, which implies $q=1/2$. In this work we fix
$q_{i}=1/2$ reducing to three the number of free parameters. In
principle, with this assumption we are losing generality but the
question is: how much $q_{i}$ can deviate from $1/2$ during large
scale structure formation? In the general relativity framework, in
some models with a constant coupling ($\delta$) between dark matter
and dark energy this condition ($q_{i}= 1/2$) is not satisfied. In
this case we have $H^{2}\propto (1+z)^{(3+\delta)}$ during matter
domination. But, what are the allowed values for $\delta$? In
\cite{guo07} it has been shown that background cosmological tests,
impose |$\delta| < 0.1$. Taking into account matter perturbations
stronger constraints on the coupling can be obtained \cite{fabris}.
Another possibility is to consider models in which matter has
pressure, such that when it dominates $q_{i}\neq 1/2$. However, if
matter perturbations are adiabatic, due to a finite speed of sound,
the mass power spectrum will present instabilities ruling out these
models unless $p=0$ ($q=1/2$) or very close to it. In principle, it
is possible to circumvent this kind of problem by assuming entropy
perturbations such that $\delta p=0$ \cite{reis}. However, in this
case the models may have problems with lensing skewness as pointed
out in \cite{reis2}. Although there are some indications that by
fixing $q_{i}=1/2$ we are not losing much in our description of the
majority of the viable models, relaxing this condition requires
further investigation, and we leave it for future work.

In the specific case $q_{i}=1/2$ we have ,
\begin{equation}
\Omega _{m\infty }=\left( 1-\frac{1}{2q_{f}}\left( 1+z_{t}\right) ^{1/\tau }\right) ^{-\tau
(1-2q_{f})}.  \label{omeff}
\end{equation}%
With the above definition, we can eliminate $z_{t}$ from equation (\ref{hq0}%
) and rewrite it as
\begin{eqnarray}
&&\left( \frac{H(z)}{H_{0}}\right) ^{2}=\left( 1+z\right) ^{3} \notag \\
 \times  &&\left( \Omega _{m\infty }^{\frac{1}{\tau (1-2q_{f})}}+(1-\Omega
_{m\infty }^{\frac{1}{\tau (1-2q_{f})}})(1+z)^{-\frac{1}{\tau
}}\right) ^{\tau (1-2q_{f})}.  \label{hq1b}
\end{eqnarray}%
The above expression for $H(z)$ in terms of $\Omega _{m\infty }$ is
very useful to make connections with models already discussed in the
literature. For instance, it is simple to verify from (\ref{hq1b})
that the parametrization (\ref{novoq}), in the special case
$q_{i}=1/2$, is related to the \textquotedblleft Modified Polytropic
Cardassian\textquotedblright\ (MPC) model \cite{gondolo}. This model
depends on three parameters: $m$
(denoted by $q$ in \cite{gondolo}), $n$ and $\Omega _{m0}$. If we identify, $%
\Omega _{m0}=\Omega _{m\infty }$, $m=1/(\tau (1-2q_{f}))$ and $n=2/3(1+q_{f})
$, it follows that the two models have the same kinematics. Note that, since $%
q_{f}<0$, the condition $n<2/3$ follows naturally. We remark that in the MPC model, $%
\Omega _{m0}$ is the present value of the matter density parameter,
while $\Omega _{m\infty }$ is defined at high redshift. These two
quantities do not necessarily have the same value in the general
case if, for instance, dark matter and dark energy are coupled
\cite{coup}. As an example, consider models with a variable coupling
between dark matter and dark energy (assumed to have constant
equation of state $w_x$), and such that
$\rho_X/\rho_m=\rho_{X0}/\rho_{m0} a^{\xi}$ \cite{dalal}. These
models can be described by Eqn. (\ref{hq1b}) if we identify
$\Omega_{m0}=\Omega _{m\infty }^{1/\tau(1-2q_f)}$, $\xi=1/\tau$ and
$w_X=-(1-2q_f)/3$. As remarked before, in our formulation it is not
necessary to make strong assumptions about the dark sector or
gravity theory. In MPC model the universe components are specified
to be matter and radiation; there is no dark energy. The
parametrization (\ref{novoq}) includes the MPC model (and the
coupling models above) as special cases.

Neglecting baryons, the quartessence Chaplygin model
$(p=-M^{4(\alpha +1)}/\rho ^{\alpha
})$ \cite{quartessence}, is obtained if we assume $q_{i}=1/2$%
, $q_{f}=-1$, identify $1/\tau =3(1+\alpha )$  and $\Omega _{m\infty
}=(1-w_{0})^{1/(1+\alpha )}$, where $w_{0}=-M^{4}/\rho _{0}^{\alpha
+1}$ is the present value of the equation of state parameter.

The conventional dark energy model with constant equation of state
($w_{X}$) is obtained if we identify $\Omega _{m\infty }=\Omega
_{m0}$ and impose the condition $-3w_{X}=1/\tau =(1-2q_{f})$ in Eqn.
(\ref{hq1b}). In particular, if $q_{f}=-1$ and $\tau =1/3$,
$\Lambda$CDM is recovered. For this model the transition redshift is
equal to $(2(1-\Omega _{m0})/$ $\Omega _{m0})^{1/3}-1$. Identifying
$\Lambda$CDM in the parameter space is very convenient; it fits
current data quite well and we should expect the \textquotedblleft
true\textquotedblright\ cosmology not to be far from this limit. We
remark that, in the framework of general relativity with
non-interacting dark matter and dark energy, if $\tau <1/3$
and $q_{f}=-1$, the dark energy component will present a transient phantom ($%
w<-1$) behavior, that could either have started in the past or in
the future ($z<0$). Models with $\tau >1/3$ are always non-phantom.

It is curious that if we apply the definition of $\tau $, given by Eqn. (\ref%
{tau}) (assuming $q_{i}=1/2$ and $q_{f}=-1$), to the flat DGP
brane-world model \cite{dgp} we obtain $\tau =1/2$, independent of
$\Omega _{m0}$. Therefore q-models with $z_{t}=(2(1-\Omega
_{m0})^{2}/$ $\Omega _{m0})^{1/3}-1$ (the DGP redshift transition)
and $\tau \approx 1/2$ are expected to be a good approximation for
flat DGP models.

\begin{figure*}[tbp]
\begin{center}
\includegraphics[width=16cm,
  angle=0]{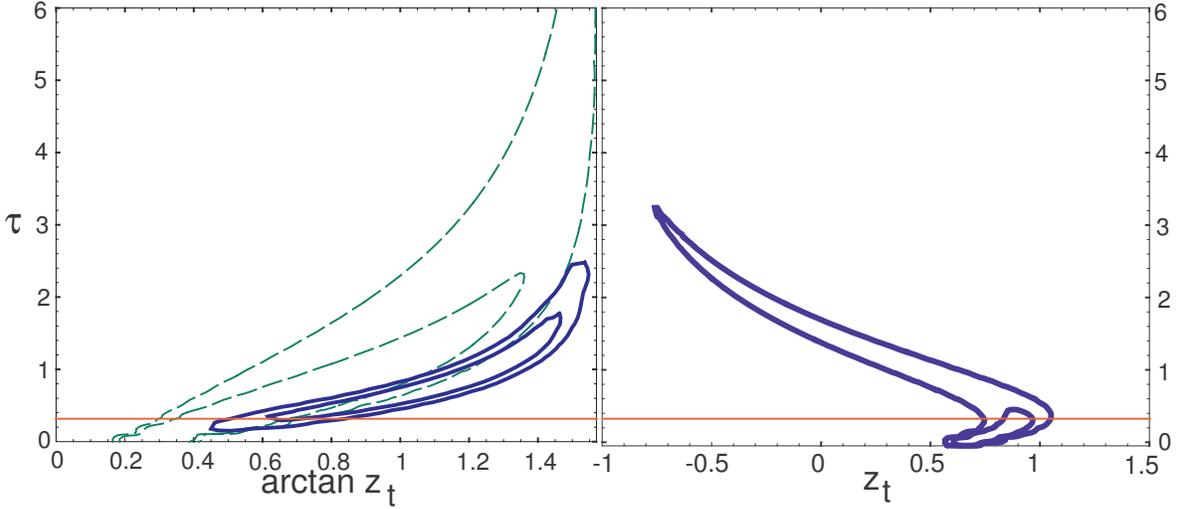}
\end{center}
\caption{{\protect {\textbf{Left} - Constraints imposed by SN Ia
observations. The contours represent $68\%$ and $95\%$ c.l. . The
green-dashed curves are from \textit{Gold182} data set and the
blue-solid ones stand for SNLS first year data set. \textbf{Right} -
Constraints imposed by $S_{k}/D_{v}$ measurements \cite{percival07}
($68\%$ and $95\%$ c.l.) as explained in the text. The red
horizontal line in both panels corresponds to flat $\Lambda$CDM
models. Notice that, for these models, higher values of $z_t$
correspond to smaller values of $\Omega_{m0}$. For both panels
$q_f=-1$ is assumed.}}} \label{fig1}
\end{figure*}

\section{Observational Constraints}

One of the main questions today in cosmology is to know if cosmic
acceleration is generated by a cosmological constant or not. The
data seem to indicate that models \textquotedblleft
close\textquotedblright\ to $\Lambda$CDM are favored. In our
analysis we first consider the special case of models that have a
final de Sitter phase ($q_{f}=-1$). In this case, the flat
$\Lambda$CDM model is more easily identified in the parameter space
allowing a simple test of the $\Lambda$CDM paradigm. The more
general case, with arbitrary $q_{f}$, will also be briefly
considered.

Assuming $q_f=-1$ we now derive constraints on the parameters $\tau
$ and $z_{t}$ by combining supernovae measurements with the ratio of
the comoving distance to the last scaterring surface,
$S_{k}(z_{ls}=1098)$, to the BAO distance scale, $D_{v}(z)$, at
$z_{BAO}=0.2$ and $z_{BAO}=0.35$, as estimated in \cite{percival07}.
In fact, the ratio $S_{k}/D_{v}$ times $z_{BAO}$ is equal to the
ratio of the CMB shift parameter ($\mathcal{R}$) \cite{shift} at
$z_{ls}$ to the BAO parameter $\mathcal{A}(z_{BAO})$
\cite{eisenstein}. This observable is appropriate for our purpose
for two reasons. First, it does not explicitly depend on the exotic
dark constituents of the universe and neither on the gravity theory.
It is essentially controlled by the function $H(z)/H_{0}$. Second,
complementarity with the SN constraints is generated because the
$S_{k}/D_{v}$ ratio and SN are sensitive to distances to objects
(events) in different redshift range; with supernovae we are
measuring distances up to $z\sim 1-2$, while $S_{k}$ depends on the
comoving distance to $z\sim 1100$.

In the SN Ia analysis we considered both, the \textit{Gold182}
\cite{gold} and the SNLS \cite{snls} samples. To determine the
likelihood of the parameters we follow the same procedure described
in these two references. In our computations, when marginalizing
over the Hubble parameter, we use a Gaussian prior such that
$h=0.72\pm 0.08$ \cite{hst}. In Fig.\ref{fig1} (left-panel),
assuming  $q_f=-1$, we display constant confidence contours (68\%
and 95\%) in the ($\arctan z_{t},\tau)$ plan allowed by SN
experiments. Notice that, for the two SN data sets $z_{t}<0$ is not
allowed at a high confidence level, indicating that a transition
occurred in the past. We remark that this is expected in
$\Lambda$CDM models (or other models) that have a non-null
transition time, but our results indicate that this is true even if
the transition is instantaneous ($\tau =0$). This conclusion also
applies if $q_f\neq-1$. Furthermore, it is also clear in
Fig.\ref{fig1} (left-panel) that current SN observations cannot
impose strong constraints on the maximum value allowed for $z_{t}$.
Since SN observations prove the universe only up to redshift $z\sim
1-2$, in a model in which the transition is slow ($\tau \gtrsim 1$),
even if $z_{t}$ is high, the distance to an object, let say, at
$z\lesssim 1$, can be similar to the distance to the same object in
another model in which $z_{t}\lesssim 1$ with a faster transition
(smaller $\tau $). This explains the shape of the SN contours. By
comparing the confidence contours for the two data sets, we observe
that those from \textit{Gold182} are shifted to lower $z_t$ with
respect to those from SNLS. We remark that even in the region of
more interest ($z_{t}\lesssim 1$), the difference between the
outcome of the two SN Ia data sets, although not so severe, exists
and is important. Similar results were obtained in
\cite{nesseris07}, which can be related to possible inhomogeneities
present in the \textit{Gold182} sample and should be further
investigated.

To obtain the constraints on the parameters from the $S_{k}/D_{v}$
test, we use a $\chi ^{2}$ statistics taking into account the
correlation matrix and the ratio $r_{s}/S_{k}$ given in
\cite{percival07}. Since we are assuming flat space we have,
$D_{v}(z_{BAO})=[z_{BAO}H^{-1}( \int_{0}^{z_{BAO}}
d\tilde{z}{H)^{2}}] ^{1/3}$ and $
S_{k}=\int_{0}^{1098}d\tilde{z}H_{e}^{-1}$. The $H_e$ term in the
definition of $S_k$ incorporates the necessity of taking into
account the contribution of radiation at very early times, which is
not included in our parametrization Eqn. (\ref{novoq}). One may
argue that by introducing radiation we are losing generality.
However, any late time modification of the standard cosmological
model should satisfy Big-Bang nucleosynthesis (BBN) constraints at
very early time. We know that radiation exists and what should be
the dependence of the Hubble parameter with redshift at very early
times (when radiation dominates) in order not to spoil BBN's
success. During this phase $H^2\propto (1+z)^{4}$ and $q=1$.
Therefore, to take radiation into account, we add the term $\Omega
_{r0}(1+z)^{4}$ to the right hand side of (\ref{hq0}) when applying
it to calculate $S_k$. If we do not consider it, we would have a
$\sim 18\%$ error in estimating ${S}_{k}$.

We marginalize the likelihood over $h$ with the same Gaussian prior
used in the supernovae analysis. In fact, $S_{k}$ is almost
independent of $h$; the dependence entering only through the
radiation term. In Fig.\ref{fig1} (right-panel) we show constant
confidence contours (68\% and 95\%) in the ($z_{t},\tau $) plan
allowed by the $S_{k}/D_{v}$ test. It is worth to be mentioned that
the shape of the $95\%$ contour is similar to contours of constant
$\Omega _{m\infty }$ and we can think that this test essentially
constrains this quantity. The same kind of behavior also appears in
flat, constant $w$ models \cite{percival07}. Furthermore, it is
clear from the figure the complementarity between the SN and this
test.
\begin{figure*}[tbp]
\begin{center}
\includegraphics[width=16cm,
  angle=0]{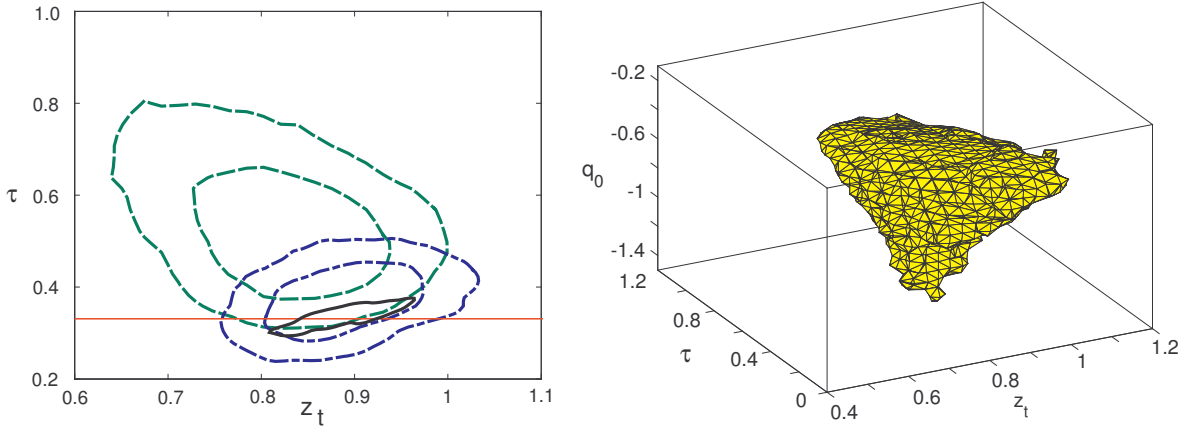}
\end{center}
\caption{{\protect {\textbf{Left} - $68\%$ and $95\%$
confidence levels imposed by the combined data sets of figure (\protect\ref%
{fig1}). The green-dashed (blue-dot-dashed) contours represent
\textit{Gold182} (SNLS)+ $S_{k}/D_{v}$. The small black-solid
contour ($95\%$ c.l.) was obtained from simulated data as explained
in text. For the figure $q_f=-1$ is assumed. \textbf{Right} - We
show, in the parameter space $(z_t,\tau,q_0)$, the $95\%$ confidence
surface for the $q_f$ general case  ($q_f \in (-\infty,0)$),
obtained using SNLS+$S_{k}/D_{v}$ data. }}} \label{fig2}
\end{figure*}

To get the combined (SN+$S_{k}/D_{v}$) results we multiply the
marginalized likelihood functions. In Fig.\ref{fig2} (left-panel) we
show the results ($68\%$ and $95\%$ c.l.) of the $S_{k}/D_{v}$ test
with the \textit{Gold182} (green dashed contours) and with the SNLS
(blue dot-dashed contours) data set. The red horizontal line in the
figure represents the $\Lambda$CDM limit ($\tau =1/3$). It shows
that this model is in good agreement with the $S_{k}/D_{v}$ + SNLS
data. After marginalizing over the extra parameter \cite{betocs} we
find for $S_{k}/D_{v}$+\textit{Gold182} (at $95.4\%$ confidence
level), $z_{t}=0.84\pm _{0.17}^{0.13}$ and $\tau =0.51\pm
_{0.17}^{0.23}$, while for $S_{k}/D_{v}$+SNLS we have,
$z_{t}=0.88\pm _{0.10}^{0.12}$ and $\tau =0.35\pm _{0.10}^{0.12}$.
For a model with $q_f=-1$, $\tau = 0.35$ and $z_t=0.88$, we obtain
from Eqn. (\ref{omeff}) that $\Omega _{m\infty }=0.23$. It is also
simple to show that the age of the universe in this particular model
(assuming $h=0.72$) is $14.0$ $Gyr$ and that cosmic acceleration
started $7.2$ $Gyr$ ago. Notice that the $\Lambda$CDM cosmology
($\tau =1/3$) is in good agreement with $S_{k}/D_{v}$ + SNLS, but
excluded at $95.4\%$ confidence level by the
$S_{k}/D_{v}$+\textit{Gold182} data. This discrepancy reveals
tension between the two SN data samples and reinforces the necessity
of better SN data to clarify the issue.  We also display in the same
figure (solid contour) what should be expected from future surveys
when combining SN Ia+$S_{k}/D_{v}$ measurements. In our Monte Carlo
simulations we used as fiducial model a flat $\Lambda$CDM model with
$\Omega _{m0}=0.23$ ($\tau =1/3,z_{t}\simeq 0.88$). For SN Ia we
considered a SNAP-like survey assuming that the intercept is known.
For the $S_{k}/D_{v}$ test we used a conservative (but somewhat
arbitrary) assumption that the uncertainties will be reduced to 2/3
of their current values. We also assumed that the correlation
coefficient would remain the same. In the figure we show the $95\%$
confidence contour.

We also analyzed the broader case with arbitrary $q_f$. Our
parametrization (\ref{novoq}) allows us to determine the present
value of $q$ ($q_0$) in terms of $z_t$, $\tau$ and $q_f$. Although
the considered data sets do not impose a lower bound for $q_f$, they
do constrain $q_0$ (we found $-1.4\lesssim q_0\lesssim -0.3$). In
Fig.\ref{fig2} (right-panel) we show, in the parameter space
$(z_t,\tau,q_0)$, the $95\%$ confidence surface for the general case
($q_f \in (-\infty,0)$), obtained using SNLS+$S_{k}/D_{v}$ data.

\section{Conclusion}
In this work, with a formulation that avoids strong assumptions
about the dark sector and/or the metric theory of gravity, we showed
that by using only SN data the transition redshift (from decelerated
to accelerated expansion) could be very large ($z_t>10$). We
demonstrated the importance of combining the SN test with the
$S_k/D_v$ test to better constrain the parameters $z_t$ and $\tau$.
We introduced the parameter $\tau$ and showed its relevance to
characterize models like flat $\Lambda$CDM, DGP and others. We
confirmed that there is a tension between \textit{Gold182} and SNLS
data sets with a quite general formulation. We also exhibited what
should be expected from future $SN + S_{k}/D_{v}$ observations and,
relaxing the condition $q_f = -1$, obtained current constraints on
the parameters $z_t$, $\tau$ and $q_0$. A more detailed analysis of
the consequences of our parametrization in the case $q_f \neq -1$ is
still necessary. For instance, our parametrization is not able to
describe, in all their redshift range, models that are now
accelerating but that decelerates again in the future
\cite{frieman}. The description of the expansion history of these
models is more complicated since it requires a second transition. It
would be interesting to investigate under what conditions it would
be possible to describe, with our parametrization, the behavior of
these kind of models for $z>0$. The case $q_i \neq 1/2$ should also
be further investigated. These issues will be discussed in
subsequent work.

\section*{\bf Acknowledgements}

We thank Maur\'{\i}cio Calv\~ao, S\'ergio Jor\'as, Miguel Quartin
and Rog\'erio Rosenfeld for useful discussions and suggestions.
EEOI, AVT and IW are partially supported by the Brazilian research
agency CNPq. RRRR is partially supported by the Brazilian research
agency CAPES.


\end{document}